\documentclass{ws-procs9x6}

\begin{document}

\title{Sensitivity of the chiral phase transition of QCD to the scalar meson
sector  \footnote{\uppercase{T}his work is supported by \uppercase{H}ungarian
  \uppercase{S}cience \uppercase{F}und (\uppercase{OTKA}) under contract 
\uppercase{N}o. 046129.}}

\author{T. Herpay}  
\address{Department of Physics of Complex Systems, \\
E{\"o}tv{\"o}s University, H-1117 Budapest, Hungary\\
E-mail: herpay@complex.elte.hu} 
\author{A. Patk{\'o}s\footnote{\uppercase{S}peaker}} 

\address{Department of Atomic Physics, E{\"o}tv{\"o}s University,\\ 
H-1117 Budapest, Hungary\\ E-mail:patkos@ludens.elte.hu} 
\author{Zs. Sz{\'e}p} 

\address{Research Group for Statistical Physics of the 
  Hungarian Academy of Sciences,\\ H-1117 Budapest, Hungary\\
E-mail:szepzs@achilles.elte.hu} 
\maketitle

\abstracts{Gribov's theory of light quark confinement implies the existence of 
two kinds of scalar bound states. 
The phase diagram of the three-flavor QCD is mapped out in
 the $(m_\pi\,-\,m_K)$--plane with help of the 
$SU_L(3)\times SU_R(3)$ linear sigma model supplemented
with the assumption that the masses of the so-called 
superbound scalars do not change
under the variation of the pion and kaon mass. The phase boundary along
the $m_\pi=m_K$ line is found in the interval $15\ \rm{MeV}<m_{crit}<25\ 
\rm{MeV}$, irrespective which  $f_0 - \sigma$ linear combination is
identified with the  pure superbound state.
}
\allowdisplaybreaks
\section{Introduction}

Volodia Gribov repeatedly has expressed in the early nineteen-eighties his view
that deconfinement in QCD 
is going to turn out to be a complete analogue of atomic
ionisation, which one would not call a phase transition.  There is
increasing evidence that the temperature driven
transition from the hadronic phase to
quark-gluon plasma is a smooth crossover indeed \cite{petreczky04}.
He based his confinement theory also on an analogy with a phenomenon
of atomic physics, e.g. supercritical binding.
This would occur for light quarks irrespective to the accurate value
of their mass \cite{gribov1}. The role of
chiral symmetry is not evidently important in this theory, and the consequences
on its spontaneous breakdown were not yet fully clarified 
\cite{dokshitzer04}. 

In this contribution to the commemoration of the 75th anniversary 
of V.N. Gribov, we exploit semi-quantitatively his ideas on 
the nature of the scalar meson sector when discussing
the chiral symmetry restoring 
quark-hadron transition in different points of the
 $m_\pi - m_K$-plane with help of a variant of the linear sigma model
($L\sigma M$).
In a recent publication \cite{herpay05} we
constructed a continuation of the parameters of this model from the
physical point to an arbitrary mass-point $m_\pi, m_K$ requiring
agreement with results of Chiral Perturbation Theory (ChPT) for the
decay constants $f_\pi, f_K$ and the trace of the squared mass matrix
in the $\eta - \eta'$ sector, $M_\eta^2$, at the tree level. In the
$\eta-\eta'$ sector the predictions of the two models for the separate
mass eigenvalues nicely coincide.  
It was emphasised that for a complete specification of 
the parameters of the model one
 needs a single extra information on the dependence of
 scalar spectra on the masses of the pseudoscalar nonet. This
information is beyond ChPT, therefore in general 
only {\it ad hoc} assumptions can be made and tested through the consequences.

According to Gribov's confinement theory in addition to normal $\bar q
q$ states also superbound states do exist, which contain quarks with {\it
  negative kinetic energy}. Some repulsive interaction between such quarks
increases their energy above zero and makes these associations
physically observable~\cite{close93}. 
These states are of smaller size and higher energy than the
normal scalar meson states, therefore it was suggested to associate
them with $f_0$ in the isoscalar and with $a_0$ in the isovector
channel. In the $SU(2)\times SU(2)$ model they form another $O(4)$
quartet in addition to $\sigma, \pi$. 
It is rather natural to assume that the masses of the superbound states
 are not sensitive to the variation of $m_\pi$.

In the three flavor case a similar ``doubling'' of the multiplet
 structure can be assumed, if the strange quark is light enough. In
 this paper we shall explore 
 the boundary of the first order transition region
around the chiral point in the three-flavor QCD \cite{pisarski84}
 taking into account the extra requirement to have scalars in the
 spectra whose mass does not vary with $m_\pi, m_K$.
The consequences
of identifying the physically observable scalars with some
mixture of the pure normal and pure superbound states will be investigated. 

  Numerical investigations of the chiral symmetry restoration were
done in the framework of lattice QCD 
and systematically improved for the 3-flavor degenerate case
$m_u=m_d=m_s\neq 0$. The initial estimate for the critical
pseudoscalar meson mass,
$m_\textrm{crit}(\textrm{diag})\approx 290$ MeV \cite{karsch01} was
seen to be reduced to $60-70$ MeV \cite{karsch03} or may be to even
further down \cite{bernard04} when finer lattices and improved lattice
actions are used. Very recently de Forcrand and Philipsen reported an
estimate $m_{crit}\approx (0.1-0.2)T_c$, which very conservatively
means $m_{crit}\approx 15-30 {\rm MeV}$ \cite{philipsen05}. 
One should be conscious of the fact, that 
it is extremely difficult to reach continuum results in this
mass regime.

Effective models (linear or non-linear sigma models,
Nambu--Jona-Lasinio model) represent another, in a sense
complementary, approach to the study of the phase structure, which one
expects to work the better the lighter quark masses are used
\cite{Meyer-Ortmanns,Lenaghan00,barducci05}. 
It is surprising that only moderate
effort was invested to date to improve the pioneering studies of the
$SU(3)~\times~SU(3)$ linear sigma model by Meyer-Ortmanns and Schaefer
\cite{Meyer-Ortmanns} which used a saddle point approximation valid in
the limit of infinite number of flavors, and derived
$m_\textrm{crit}(\textrm{diag})\lesssim 51$ MeV. An extension of their
work to unequal pion and kaon masses was achieved by 
C. Schmidt~\cite{schmidt03}. He found $m_\textrm{crit}(\textrm{diag})=47$ MeV 
and a phase boundary approaching the $m_K$-axis rather sharply.   
The phase boundary was
calculated also by Lenaghan~\cite{lenaghan01} using the
Hartree-approximation to the effective potential derived in
CJT-formalism.  For the complete determination of the couplings of the
three-flavor chiral meson model he fixed the $T=0$ mass of the
$\sigma$ particle in addition to the experimental mass spectra
 of the pseudoscalar
sector. The emerging phase boundary is rather sensitive to this mass.
 The estimate for $m_\textrm{crit}(\textrm{diag})$
which one can extract from Fig.~3 of \cite{lenaghan01} for
$m_\sigma=900$ MeV is compatible with \cite{Meyer-Ortmanns,schmidt03}.
 
Our method of parametrisation and solution of $L\sigma M$ was
described in detail in \cite{herpay05}. Therefore here we shall
only  review the set of the equations to be solved for
the determination of the transition point. The changes arising from
the implementation of the insensitivity of Gribov's superbound scalars
to the variation of the fundamental masses will be emphasised. We
conclude by giving the most characteristic features of the phase
boundary.

\section{$L\sigma M$ parametrisation consistent with ChPT} 

We outline first, how one obtains with ChPT the $m_\pi, m_K$-dependence of the
masses and decay constants in the pseudoscalar sector.
The dependence of the pion and kaon masses as well as of their decay
constants on the quark masses were determined in
\cite{Gasser85} for the $SU(3)\times SU(3)$ nonlinear sigma model:

\begin{eqnarray}
m_\pi^2&=&2{A}\left[1+\frac{1}{f^2}\bigg(\mu_\pi-\frac{1}{3}\mu_\eta
+16{A}(2L_8-L_5)+\right.\nonumber\\&&\left.
+16{A(2+q)}(2L_6-L_4)\bigg)\right],\\
m_K^2&=&{A(1+q)}\left[1+\frac{1}{f^2}\bigg(\frac{2}{3}\mu_\eta
+8{A(1+q)}(2L_8-L_5)\nonumber\right.\\&&\left.
+16{A(2+q)}(2L_6-L_4)\bigg)\right],\\
f_\pi&=&f\left[1+\frac{1}{f^2}\left(-2\mu_\pi-\mu_K+8{A}L_5+
8{A(2+q)}L_4\right)
\right],\\
f_K&=&f\left[1-\frac{1}{f^2}\Bigl(\frac{3}{4}(\mu_\pi+\mu_\eta+2\mu_K)-
4{A(1+q)}L_5-8{A(2+q)}L_4\Bigr)\right].
\end{eqnarray}
Here $A,q$ are related to the quark masses, $f$ is the coupling of the
non-linear sigma model. $\mu_{PS}$ defines the
so-called chiral logarithm for each pseudoscalar meson (PS)
proportional to $\ln (m_{PS}/M_0)$ with $M_0=4\pi f$. The determination
of the low energy chiral constants $L_i$ is discussed in depth in
\cite{herpay05}. 

One inverts the first two equations with $ O(f^{-2})$ accuracy
and finds the following $m_\pi, m_K$-dependence for the decay
constants:
\begin{eqnarray}
\label{decayconst}
f_\pi&=&f\left[1-\frac{1}{f^2}(2\mu_\pi+\mu_K-4m_\pi^2(L_4+L_5)-
8m_K^2L_4)\right],
\\
f_K&=&f\left[1-\frac{1}{f^2}\left(\frac{3}{4}(\mu_\pi+\mu_\eta+2\mu_K)-
4m_\pi^2L_4-4m_K^2(L_5+2L_4)\right)\right].\nonumber
\end{eqnarray}
The extension to the $U(3)\times U(3)$ ChPT is somewhat more
complicated. It was worked out in
\cite{Herrera98,Herrera97,Borasoy01,Beisert01} and allows the
determination of $m_\eta(m_\pi, m_K), m_{\eta'}(m_\pi, m_K)$. For the
parametrisation of $L\sigma M$  one more independent relation can
be obtained from the mixing $\eta -\eta'$ sector, for which in 
\cite{herpay05} we have
chosen the trace of the $2\times 2$ squared mass matrix, denoted by
 $M_\eta^2\equiv m_\eta^2+m_{\eta'}^2$:
\begin{eqnarray}
\label{Meta-extrap}
{M_\eta^2}&=&2m_K^2-3v_0^{(2)}+2(2m_K^2+m_\pi^2)
(3v_2^{(2)} - v_3^{(1)}) \\
&+&\frac{1}{f^2}\Big[8v_0^{(2)}
(2m_K^2+m_\pi^2)(L_5+3L_4)+m_\pi^2(\mu_\eta-3\mu_\pi )
-4m_K^2\mu_\eta\nonumber \\
&+&\frac{16}{3}(6L_8-3L_5+8L_7)(m_\pi^2-m_K^2)^2\nonumber\\
&+&\frac{32}{3}L_6(m_\pi^4-2m_K^4+m_K^2m_\pi^2)+
\frac{16}{3}L_7(m_\pi^2+2m_K^2)^2 \Big]. \nonumber 
\end{eqnarray}
Also the constants $v_i^{(j)}$ were determined in \cite{herpay05} from
the $T=0$ properties of the Goldstone particles.

Next, we turn to the problem of parameterising $L\sigma M$ to provide
spectra the closest
possible to ChPT. The Lagrangian of the $SU_L(3)~\times~SU_R(3)$ symmetric 
linear sigma model 
with explicit symmetry breaking terms is given \cite{Haymaker73} by
\begin{eqnarray}
L(M)&=&\frac{1}{2}{\rm tr}(\partial_\mu M^{\dag} \partial^\mu M+\mu_0^2
M^{\dag} M)-f_1
\left( {\rm tr}(M^{\dag} M)\right)^2\\
&-&f_2  {\rm tr}(M^{\dag}M)^2
-g\left(\det(M)+\det(M^{\dag})\right)+\epsilon_0\sigma_0+\epsilon_8 
\sigma_8,\nonumber
\label{Lagrangian}
\end{eqnarray}
where $M$ is a complex 3$\times$3 matrix, defined by the $\sigma_i$ scalar
and $\pi_i$ pseudoscalar fields 
$\displaystyle
M:=\frac{1}{\sqrt{2}}\sum_{i=0}^{8}(\sigma_i+i\pi_i)\lambda_i$,
with  $\lambda_i\,:\,\,\,i=1\ldots 8$ the Gell-Mann matrices and  
$\lambda_0:=\sqrt{\frac{2}{3}} {\bf 1}.$
The last two terms of (\ref{Lagrangian}) break the symmetry
explicitly, the possible isospin breaking term $\epsilon_3\sigma_3$ is
not considered. 

A detailed analysis of the symmetry breaking patterns
which might occur in the system described by this Lagrangian can be
found in \cite{Lenaghan00}. The field expectation values
 $\langle \sigma_0\rangle, \langle\sigma_8\rangle$ both
contain strange ($\equiv y$) and non-strange ($\equiv x$) components:
\begin{equation}
x=(\sqrt{2}\langle\sigma_0\rangle+\langle\sigma_8\rangle)
/\sqrt{3}, \qquad y=(\langle\sigma_0\rangle-\sqrt{2}
\langle\sigma_8\rangle)/\sqrt{3}.
\end{equation}

With help of the PCAC relations and the tree level mass formulae (see 
Table \ref{Tab:masses}, where the $x-y$ basis is used instead of $0-8$)
the following expressions can be derived for the couplings of $L\sigma
M$: 
\begin{eqnarray}
 x&=&f_\pi, \qquad
 y=\left(2f_K-f_\pi\right)/\sqrt{2} \, ,\\
{f_2} &=&
\frac{(6f_K-3f_\pi)m_K^2-(2f_K+f_\pi)m_\pi^2-2(f_K-f_\pi){M}_\eta^2}
{4(f_K-f_\pi)(8f_K^2-8f_K f_\pi+3f_\pi^2)}, \nonumber\\
{g}&=& \frac{2f_K
  m_K^2+2(f_K-f_\pi)m_\pi^2-(2f_K-f_\pi){M}_\eta^2}{\sqrt{2}
(8f_K^2-8f_K f_\pi+3f_\pi^2)}  , \nonumber\\ 
M^2&\equiv& -\mu_0^2+4f_1(x^2+y^2)\nonumber\\
&=&\frac{1}{2}M_\eta^2+\frac{g}{\sqrt{2}}(2f_K-f_\pi)-
2 f_2 [(f_\pi-f_K)^2+ f_K^2].\nonumber
\end{eqnarray}
The sources $\epsilon_x=(\sqrt{2}\epsilon_0+\epsilon_8)/\sqrt{3}, 
\epsilon_y=(\epsilon_0-\sqrt{2}\epsilon_8)/\sqrt{3}$,
 which explicitly break chiral
symmetry are determined with help of the Gell-Mann--Oakes--Renner
relations:
\begin{equation}
\epsilon_x=m_\pi^2 x,  \quad \quad
\epsilon_y=\frac{\sqrt{2}}{2}
(m_K^2-m_\pi^2)x+m_K^2y \label{ae1}.
\end{equation}
\begin{table}
\tbl{Squared masses of the pseudoscalar boson nonet and their parity
  partners
\label{Tab:masses}}
{\begin{tabular}{|l|}
\hline
$m^2_\pi\,\,\,\,\,=-\mu_0^2+2(2f_1+f_2)x^2+4f_1y^2+2gy$ \\
$m^2_{a_0}\,\,=-\mu_0^2+2(2f_1+3f_2)x^2+4f_1y^2-2gy$ \\\hline\hline
$m^2_K\,\,\,=-\mu_0^2+2(2f_1+f_2)(x^2+y^2)+2f_2y^2-\sqrt{2}x(2f_2y-g)$\\
$ m^2_{\kappa}\,\,\,\,\,=-\mu_0^2+2(2f_1+f_2)(x^2+y^2)+2f_2y^2+
\sqrt{2}x(2f_2y-g)$ \\\hline\hline
$m^2_{\eta_{xx}}=-\mu_0^2+2(2f_1+f_2)x^2+4f_1y^2-2gy$ \\
$ m^2_{\eta_{yy}}=-\mu_0^2+4f_1x^2+4(f_1+f_2)y^2$ \\
$m^2_{\eta_{xy}}=-2gx$ \\    
$m^2_{\sigma_{xx}}=-
\mu_0^2+6(2f_1+f_2)x^2+4f_1y^2+2gy$ \\   
$m^2_{\sigma_{yy}}=- 
\mu_0^2+4f_1x^2+12(f_1+f_2)y^2$ \\
$ m^2_{\sigma_{xy}}=8f_1xy+2gx $ \\
\hline
\end{tabular}}
\begin{tabnote}
The expressions of the squared masses of  
parity partners having the same isospin and hypercharge
appear in one block. Different isomultiplets are separated
by double lines. In the lowest big block the matrix elements of the 
 mixing in the $\eta - \eta'$ and $\sigma - f_0$ sectors are given
 in the $x-y$ base.
\end{tabnote}
\end{table}

\begin{figure}[htbp]
\centering {\includegraphics[width=0.75\textwidth, 
keepaspectratio]{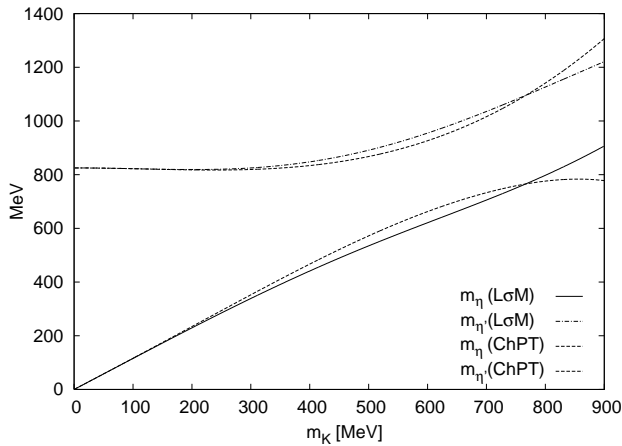}  }
\caption{The tree--level kaon mass dependence of $m_\eta$ and $m_{\eta^{'}}$ 
for $m_\pi=0$. The labels refer to the results of ChPT and the predictions
 of linear sigma model (L$\sigma$M), respectively.}
\label{fig:m_eta}
\end{figure}

The requirement of the 
agreement of $L\sigma M$ with ChPT is fulfilled  when the
$(m_\pi, m_K)$-dependence of the couplings
$x,y,f_2,g,M^2$ is determined with help of
Eqs.(\ref{decayconst},\ref{Meta-extrap}). 
Before the procedure just
described was first proposed in Ref.\cite{herpay05} $m_\pi-m_K$ mass
tuning was taken into account only in (\ref{ae1}). The
quality of this not fully complete parametrisation (e.g. only the 
combination $M^2$ of $f_1$ and $\mu_0^2$ is determined at this stage) 
can be assessed by comparing the separate the $m_\pi - m_K$ mass dependence of 
$m_\eta$ and $m_{\eta'}$ obtained in the present parametrisation of
$L\sigma M$ with the predictions of ChPT. In Fig. \ref{fig:m_eta} the
comparison is done for $m_\pi=0$ and the agreement is fairly good up
to $m_K\approx 800$MeV.

 The combination ${M}^2$ of $f_1$ and $\mu_0^2$ can be split up only
 by making use of the
expression of the admixed scalars, therefore the use of one characteristics
of the mixed scalar spectra is unavoidable \cite{Tornqvist99}. 

For the separate determination of $\mu_0^2$ and $f_1$ we have fixed
first the mass of the $f_0$ meson. This meson mass is identified with
the heavier eigenvalue calculated in the mixing scalar subspace.
Since there is the possibility of
mixing between the normal and the superbound multiplets we have chosen
as a second possibility also fixing the trace
$m_\sigma^2+m_{f_0}^2=m_{\sigma_{xx}}^2+m_{\sigma_{yy}}^2$ (see Table
\ref{Tab:masses}) 
in the squared mass subspace of admixed scalars. 
In order to test the sensitivity of the results to the scalar masses
(which are not very accurately known) we have performed the
calculations for both alternatives with several numerical values.

\section{Quasiparticle thermodynamics and phase diagram 
\label{sec:termo}}

The renormalised set of the equations of state and of the
selfconsistent pion propagator determine the temperature dependence of
the vacuum expectation values $x$ and $y$. The scheme of the
perturbation theory applied here
agrees with the Optimal Perturbation Theory of Chiku and
Hatsuda \cite{chiku}, which was renormalised using the approach of
 \cite{jako05}.

The tree level mass of $\pi$ involves now the thermal mass parameter:
\begin{equation}
m_\pi^2=M^2(T)+2(2f_1+f_2)x^2+4f_1y^2+2gy,
\label{resum-pion}
\end{equation}
and all other meson masses to be used in the tadpole integrals below
agree with the formulas appearing in Table~\ref{Tab:masses} with the
replacement $-\mu^2_0\rightarrow M^2(T)$.  If all quantum corrections
are condensed into $M^2(T)$, then the tree--level masses of other
mesons are expressible through the mass of the pion.  One might expect
that the pion has the lowest mass and therefore for $M^2(T) > 0$ these
squared masses are all positive, which is not the case when
$-\mu_0^2<0$ is used in the propagators.  We define a physical region
of $x$ and $y$ where all tree-level mass squares are positive, and
thus the one-loop contribution of the meson fluctuations to EoS is
real. This region is most severely restricted by the mass of 
$\sigma$, which strongly decreases near the phase transition. We will
restrict our attention to the solution of the EoS's in the physical region.

For the determination of the thermal mass we use the Schwinger--Dyson
equation for the inverse pion propagator at zero external momentum.
At one-loop it  receives the contribution $\Pi(M(T),p=0)$, which is the
self-energy function of the pion at zero external momentum, plus the
counterterm contribution $-\mu_0^2-M^2(T)$. We apply the principle of 
minimal sensitivity (PMS) \cite{chiku}, that is we require that the pion 
mass be given by its tree-level expression:
\begin{equation}
\label{Eq:M_T_gap}
\Pi(M(T),p=0)-\mu_0^2-M^2(T)=0.
\end{equation}
$\Pi(M(T),p)$ itself is a linear combination of the tadpole and bubble
diagrams (the latter not included in the treatment of
\cite{Lenaghan00}), with coefficients derived with help of the 4-point
and 3-point couplings among mass eigenvalue fields. 

The self-energy can be represented as a linear
combination of tadpole integrals, which gives when substituted into 
Eq.~(\ref{Eq:M_T_gap}):
\begin{equation}
0=-M^2(T)-\mu_0^2+
\sum^{\alpha=\sigma,\,\pi}_{i=\pi,\,K,\,\eta,\eta^{'}} c^\pi_{\alpha_i
} I(m_{\alpha_i}(T),T) \, . \label{gap}
\end{equation}
Here $c^\pi_{\alpha_i}$ are the weights of the renormalised
tadpole contributions
evaluated with different mass eigenstate mesons $\alpha_i=\sigma_i, \pi_i$.
The integrals over the corresponding propagators are
evaluated with effective tree-level masses where $M^2(T)$ replaces
$-\mu_0^2$. In this way (\ref{gap})
is actually a gap equation which determines the thermal mass
parameter, $M^2(T)$. With help of Eq. (\ref{resum-pion}) this
equation can be also understood as a gap equation for the pion mass
(the pion mass is present also in the expressions of
$I(m_{\alpha_i},T)$ through $m_{\alpha_i}$!):
\begin{equation}
m_\pi^2=-\mu_0^2+2(2f_1+f_2)x^2+4f_1y^2+2gy+\sum^{\alpha=
\sigma,\,\pi}_{i=\pi,\,K,\,\eta,\eta^{'}} c^\pi_{\alpha_i
} I(m_{\alpha_i}(T),T) \, .
\label{pigap}
\end{equation}

For the determination of the order parameters $x,y$ we
 solved the two renormalised equations of state using the solution of
 the gap equation in the propagator masses: 
\begin{eqnarray}
\label{xeq}
&-\epsilon_x-\mu_0^2 x+2gxy+4f_1xy^2+2(2f_1+f_2)x^3\\
&+\sum^{\alpha=\sigma ,\pi}_{i=\pi,\,K,\,\eta,\eta^{'}}  J_i t^x_{\alpha_i
} I(m_{\alpha_i}(T),T)=0 \,,\nonumber\\
\label{egy2}
&-\epsilon_y-\mu_0^2 y+gx^2+4f_1x^2y+4(f_1+f_2)y^3\\
&+\sum^{\alpha=\sigma ,\pi}_{i=\pi,\,K,\,\eta,\eta^{'}}  J_i t^y_{\alpha_i
} I(m_{\alpha_i}(T),T)=0\,,\nonumber 
\end{eqnarray} 
The quantities $t^x_{\alpha_i}$ and $t^y_{\alpha_i}$ give the corresponding
weights. $J_i$ is the isospin multiplicity factor: 
$J_\pi=3$, $J_K=4$, and  $J_{\eta,\eta'}=1$.  The coefficients $c$ and
$t$ were listed in \cite{herpay05}. One finds
$c^\pi_{\alpha_i}=J_i t^x_{\alpha_i}/x$, which ensures that the
solution for the mass of the pion obeys Goldstone's theorem.  

The solution of Eqs. (\ref{pigap}), (\ref{xeq}), 
(\ref{egy2}) for given $m_\pi, m_K$  allows to
establish the nature of the temperature driven transitions.
First order transitions are signalled by multivaluedness in the
temperature evolution of both the non-strange and strange condensates.  
For large values of the kaon mass, we claim
that the phase transition is driven by the variation of the
non-strange condensate, since each of the multiple solutions of
the strange condensate are very close to each other, and all stay at
high values. Subsequent decrease of the strange condensate at higher
temperature displays only a crossover.

\begin{figure}[htbp]
\vspace*{-10pt}
\centering {\includegraphics[width=0.75\textwidth, 
keepaspectratio]{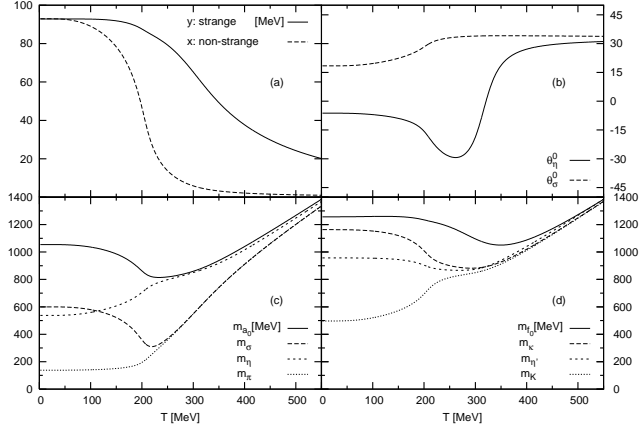}}
\caption{The temperature dependence  in the physical point of: 
(a)~the non-strange (x) and strange 
(y) condensates; (b)~the pseudoscalar ($\theta_\eta$) and scalar 
($\theta_\sigma$) mixing angles (in the (0-8) basis); 
(c)~the mass of the chiral 
partners ($\pi,\sigma$) and ($a_0,\eta$); (d)~the mass of
$f_0,\kappa,\eta',K$ mesons.} 
\label{fig:phys_vev_masses}
\end{figure}

\begin{figure}[htbp]
\centerline{\includegraphics[width=0.75\textwidth,
keepaspectratio]{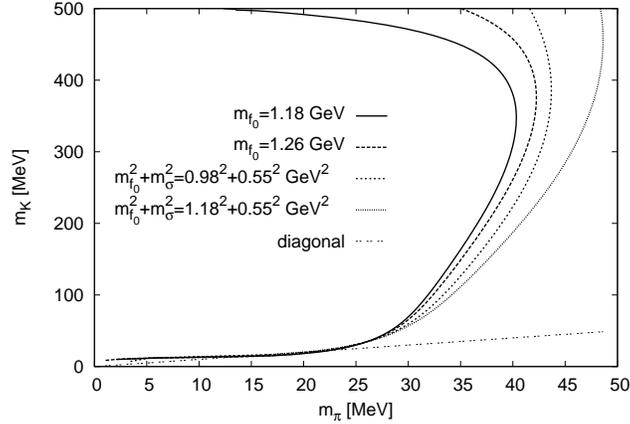}}
\caption{Variation of the boundary of first order transitions near and
  above the
$m_\pi=m_K$ diagonal (note the rather different units on the two axes!)}
\label{fig:boundary}
\end{figure}

In the physical point the system exhibits clear crossover as can be
seen in Fig.\ref{fig:phys_vev_masses}.
Let us discuss next what happens near the $m_\pi=m_K$ diagonal. In
Fig.\ref{fig:boundary} the gradual 
deformation of the boundary line is shown when
four different $(m_\pi, m_K)$-independent conditions (listed on the figure) 
are imposed on the spectra of the admixed scalars. 
The boundary reaches the diagonal in the range
$m_{crit}\approx 20$ MeV independently of the condition imposed. When
moving away from the diagonal for
$m_K> m_\pi$ the larger is the scalar mass scale the farther the
boundary goes away from the $m_\pi=0$ axis. Therefore with this method
one cannot make any definite statement
on the location of the tricritical point on the $m_K$-axis defined as
the point where the transition changes  from a
discontinuous nature into a continuous one.

In conclusion of this study we find that the assumption of the
$m_\pi-m_K$ independence of the mass of the heavier
isoscalar-scalar member of the nonet, suggested by Gribov's
confinement picture leads to a rather unique conclusion, that the
$m_\pi=m_K$ diagonal crosses the critical curve of the chiral
phase transitions of the 3-flavor QCD at a rather low value: 
$15\ \rm{MeV}\leq m_{crit}\leq 25\ \rm{MeV}$.

\section*{Acknowledgement}
We thank the organisers of the Gribov75 conference for the
opportunity to discover again the actual interest of the confinement
theory of V.N. Gribov.

\end{document}